\documentstyle[twocolumn,prd,aps]{revtex}
\begin{document}
\draft
\title{Propagation of Cool Pions}
\author{Robert D. Pisarski and Michel Tytgat}
\address{Department of Physics, Brookhaven National Laboratory, 
Upton, New York 11973-5000, USA}                                        
\date{\today}
\maketitle
\begin{abstract}
For an exact chiral symmetry which is spontaneously
broken at zero temperature, we show that at nonzero temperature,
generally pions travel at {\it less} 
than the speed of light.  
This effect first appears at next to leading
order in an expansion about low temperature.
When the chiral symmetry is approximate we obtain
two formulas, like that of Gell-Mann, Oakes, and Renner,
for the static and dynamic pion masses.
\end{abstract}
\pacs{BNL Preprint BNL-PT-961, February, 1996.}
\begin{narrowtext}

Pions are light because they 
are (almost) Goldstone bosons:
in $QCD$, quarks have an (approximate) chiral symmetry of 
$SU(2)_\ell \times SU(2)_r$ which 
is spontaneously broken to the usual isospin symmetry
of $SU(2)_V$ by the dynamical generation of a quark
condensate \cite{gold,donoghue,sv}. 
Notably, the pion mass squared is proportional to
the up and down quark masses through the formula of
Gell-Mann, Oakes, and Renner \cite{gmor,dashen}.

In this paper we consider how pions propagate in a thermal
bath \cite{rpmt}; similar results should also hold for pions propagating
in a Fermi sea of nucleons.  In the limit of exact chiral symmetry,
pions are true Goldstone modes and so massless.  At zero temperature
relativistic invariance then requires pions to travel at the speed of
light.  Our basic point is elementary: since
the presence of a medium
provides a privileged rest frame, relativistic invariance 
no longer applies, and so typically pions travel at {\it less} than
the speed of light.
We also derive how the formula of
Gell-Mann, Oakes, and Renner generalizes to
nonzero temperature \cite{gmort}.  Because the pion's
velocity is less than $c$,
in a medium the pion dispersion relation, as a function
of momentum, is ``flattened'' from that at zero temperature.
Such a flattening has been found in a wide variety of 
models \cite{pioncon,shuryak,nucl}, due apparently to 
the detailed dynamics.  Our results show
that at least some of the flattening arises on {\it very}
general grounds, as a consequence of chiral symmetry breaking
in a medium.

While we speak of pions throughout, our conclusions
apply to Goldstone bosons in any
system which is relativistically invariant at zero temperature.
Indeed, our results for the changes in the pion dispersion
relation have exact analogies with spin waves
in antiferromagnets \cite{anti,leutnon}.  We think that our manner
of derivation --- in terms of the pion decay constants ---
is novel and illuminating.  
The detailed calculations which we perform to demonstrate
this effect in the linear sigma model
extend and complement previous results by
Itoyama and Mueller \cite{im}.

We begin with a heuristic derivation,
in the limit of exact chiral symmetry.  At zero temperature,
the matrix element of the axial vector current, $A^\mu_{a}$, 
sandwiched between the vacuum and a pion of momentum 
$P^\mu = (p^0,\vec{p})$ is:
\begin{equation}
\langle 0 \vert A^\mu_{a} \vert \pi^b(P) \rangle 
= i f_\pi \delta^{ab} P^\mu \; ,
\label{ea}
\end{equation}
with $a$ and $b$ isospin indices.
The pion decay constant $f_\pi \sim 93 MeV$; whenever we
write $f_\pi$, we mean its value at zero temperature.

At nonzero temperature, because of the presence of the
medium, we expect that there are {\it two} distinct
pion decay constants, one for the timelike component of the
current, $f_\pi^t$,
$$
\langle 0 \vert A^0_{a} \vert \pi^b(P) \rangle_T 
= i f^t_\pi \delta^{ab} p^0 \; ,
$$
and one for the spatial, $f_\pi^s$,
\begin{equation}
\label{ftfs}
\langle 0 \vert A^i_{a} \vert \pi^b(P) \rangle_T 
= i f^s_\pi \delta^{ab} p^i \; .
\label{eb}
\end{equation}
Both matrix elements are computed at a temperature $T$
in the imaginary time formalism.
Implicitly the timelike component
of the momentum, $p^0$, is analytically continued from
euclidean values (pions, as bosonic fields, have
$p^0 = 2 \pi n T$ for integral $n$)
to Minkowski values, $p^0 = - i \omega + 0^+$.  
In (\ref{eb}), $f_\pi^t$ and $f_\pi^s$ are defined
about zero momentum, $\omega$ and $p \rightarrow 0$.

The possibility of two distinct pion decay constants
is familiar from nonrelativistic systems, such
as discussed by Leutwyler \cite{leutnon}; in this context
it was recognized previously by Kirchbach and Riska \cite{kr} and
by Thorsson and Wirzba \cite{gmort}.

By assumption the chiral symmetry is exact, and only
broken spontaneously by the vacuum.  Consequently,
while the axial current acts nontrivially on the
vacuum, it nevertheless is conserved on the pion mass
shell.  At zero temperature this is trivial:
the divergence of the matrix element in (\ref{ea})
is $\langle 0 \vert \partial_\mu A^\mu \vert \pi \rangle 
\sim f_\pi P^2$,
which vanishes when $P^2 = - \omega^2 + p^2 = 0$,
as expected for a massless, relativistically invariant field.

At nonzero temperature, however, the condition that the
axial current is conserved on the pion mass shell leads to 
interesting restrictions on the two pion decay constants.
The divergence of the axial current in (\ref{eb}) vanishes when
\begin{equation}
f_\pi^t \, p_0^2 + f_\pi^s \, p^2 = 0|_{\pi \; mass \; shell} 
\label{ec}
\end{equation}
At nonzero temperature each pion decay constant, $f_\pi^t$ and $f_\pi^s$,
has a real and an imaginary part.  
The pion mass shell then lies in the complex plane, at
$p^0 = - i \omega - \gamma$.  Equating the real parts of
(\ref{ec}) gives
\begin{equation}
\omega^2
= v^2 p^2 \approx
\frac{\mbox{Re} \, f_\pi^s}{\mbox{Re} \, f_\pi^t} \; p^2 \; .
\label{ed}
\end{equation}
The requirement that pions travel at less
than (or equal to) the speed of light, $v \leq 1$, implies 
$\mbox{Re} \, f_\pi^s \leq \mbox{Re} \, f_\pi^t$.
To obtain this, we assume that the imaginary parts can be
neglected relative to the real parts,
\begin{equation}
\mbox{Im} \, f_\pi^{t,s} \ll \mbox{\mbox{Re}}\, f_\pi^{t,s} \; .
\label{eea}
\end{equation}
Physically, $v \leq 1$
is most familiar: pions move through a medium as if it has an index
of refraction greater than or equal to one.

The imaginary part of the mass shell is given by
\begin{equation}
\gamma \approx \frac{1}{2 \omega \, \mbox{\mbox{Re}}\, f_\pi^t}
\left( + \mbox{Im} \, f^t_\pi \; \omega^2 - \mbox{Im} 
\, f^s_\pi \; p^2 \right) \geq 0 \; .
\label{eab}
\end{equation}
The requirement that pions are damped, and not anti damped,
fixes $\gamma$ to be semi positive definite; using (\ref{ed}),
this then constrains the real and imaginary
parts of $f^t_\pi$ and $f^s_\pi$.

Our analysis only applies to
``cool'' pions, where the components of the pion momenta,
$\omega$ and $p$, are small relative to the 
real parts of $f^t_\pi$ and $f^s_\pi$.
If the chiral phase transition is of second order at $T=T_\chi$,
then as $T \rightarrow T_\chi^-$,
$f^t_\pi(T)$ and  $f^s_\pi(T) \rightarrow 0$ \cite{fpi},
and the region in which cool pions dominate shrinks to zero.
About $T_\chi$, over large distances
the behavior of pions (and the $\sigma$ meson)
is controlled by an $O(4)$ critical point,
as appropriate for two massless flavors.  

Assuming that the imaginary parts of $f^t_\pi$ and $f^s_\pi$ are
nonzero at zero momentum, from (\ref{eab}) the damping rate vanishes
linearly about zero momentum, $\gamma \sim p$ as $p \rightarrow 0$.
This is consistent with Goldstone's theorem \cite{gold}: at zero
momentum, the complete inverse pion propagator must vanish, including
both the real and the imaginary parts.  If $\gamma \sim p$, then the
imaginary part of the pion self energy, $Im \Pi(P)$ in (10), and so
the complete inverse pion propagator $\Delta^{-1}(P)$, vanishes $\sim
p^2$ as $p \rightarrow 0$. This implies that even when pions are
damped, about zero momentum they still dominate the correlation
functions of axial vector currents.

In a nonlinear sigma model, the first contribution to the
damping rate appears at two loop 
order \cite{non2,schenka,schenkb}.  Using a virial
expansion, such as (2.4) of ref. \cite{schenka}, we
estimate that about zero momentum in the chiral limit,
$\gamma \sim p (T^4/f_\pi^4)$.  In the linear sigma model
considered below, the damping rate vanishes
exponentially, (\ref{fa}), but this is special to the
kinematics at one loop order in this model \cite{antidamp}.

To make our conclusions rigorous, and to extend them to an approximate
chiral symmetry, we follow
Shore and Veneziano \cite{sv} by
using a chiral Ward identity of $QCD$.
Take two flavors of quarks, each with a (current) quark mass
$=m$.  A chiral Ward identity between the form factors and the
propagators of the quark composite operator $\phi^a_5 \,\hat=\,i
\overline{q} t^a \gamma_5 q$ is \cite{sv}
\begin{equation}
\label{wi}
\partial_\mu \langle 0 \vert 
A^\mu_a \vert \phi_5^b \rangle_T 
+ \langle \bar q q\rangle_T \, \langle 0 \vert {\cal T}^\ast 
\phi_5^a \phi_5^b \vert 0\rangle_T^{-1} = 2 m\, \delta^{a b}\; , 
\end{equation}
where $\langle \bar q q\rangle_T$ is the quark condensate
and $\langle 0 \vert {\cal T}^\ast 
\phi_5^a \phi_5^b \vert 0\rangle_T^{-1}$ the inverse
propagator for the $\phi_5^a$ field.
As usual, this chiral Ward identity has the same
structure as at zero temperature,
except that now thermal expectation values enter.
Assume that  $\phi_5$ is directly proportional to
the pion field, 
\begin{equation}
\label{norm}
\pi^a = b \phi_5^a \; .
\label{ef}
\end{equation}
The normalization constant ``$b$'' is a function of 
both temperature and momentum.  The temperature
dependence follows from our analysis, while we
neglect any momentum dependence.
As discussed by 
Shore and Veneziano \cite{sv}, dropping this
momentum dependence is equivalent to the usual
assumptions which give the partial conservation of
the axial vector current.

Using (\ref{ftfs}) and (\ref{norm}) in (\ref{wi}),
the chiral Ward identity becomes
\begin{equation}
- b\left( f_\pi^t p^2_0 + f_\pi^s p^2 \right) 
+ b^2 \langle \overline{q} q\rangle_T \Delta_\pi^{-1}(P) = 2 m\; .
\label{eg}
\end{equation}
For the inverse pion propagator $\Delta_\pi^{-1}(P)$ we take 
\begin{equation}
\Delta_\pi^{-1}(P) = p_0^2 + v^2 p^2 + m_\pi^2 - i \, \mbox{Im} \Pi(P) \; .
\label{eh}
\end{equation}
Similar forms of the pion propagator have appeared previously
\cite{pioncon,shuryak,nucl}; for us this form is motivated by
the need to satisfy the chiral Ward identity.

Ref. \cite{sv} requires that the pion field is canonically
normalized, so the
coefficient of $p_0^2$ 
in the pion propagator must be unity.
We allow
for a pion velocity which is less than one by introducing
the velocity ``$v$''.  Since the quark mass
$m\neq0$, we introduce a pion mass, $m_\pi$.  Lastly, we introduce
an imaginary part of the pion self energy, $\mbox{Im} \Pi(P)$, which
is a function of momentum.
This form of the propogator should be valid in an expansion
about zero momentum.  

The chiral Ward identity shows that the assumption
used to derive (\ref{ec}) is correct: 
in the chiral limit, $m=0$, the divergence of the
axial current vanishes on the pion mass shell,
as defined by the condition
$\Delta_{\pi}^{-1}(P) = 0$.  Since the
chiral Ward identity holds for arbitrary (small) momentum, however,
we can derive several identities by matching the coefficients
of $p_0^2$, $p^2$, and $1$, 
for both the real and imaginary parts.

Equating the terms $\sim p_0^2$ fixes the constant of 
proportionality between the quark operator and the pion field
to be
\begin{equation}
b = \frac{\mbox{Re} \, f_\pi^t}{\langle\overline{q} q\rangle_T} \; .
\label{ei}
\end{equation}
Since both terms
on the right hand side of (\ref{ei}) change with temperature,
so does the factor ``$b$''.
Matching the terms $\sim p^2$ fixes the velocity as in
(\ref{ed}).
Lastly, matching the imaginary parts in (\ref{eg}) gives
$ \mbox{Im} \Pi(P) = 2 \omega \gamma$, with $\gamma$ as in (\ref{eab}).

Away from the chiral limit, we match the real parts 
at zero momentum, $p_0 = p = 0$, to obtain
the generalization of the relation
of Gell-Mann, Oakes, and Renner to nonzero temperature:
\begin{equation}
m^2_\pi = \frac{2 m \langle \overline{q} q\rangle_T}
{(\mbox{Re} \, f_\pi^t)^2} \; .
\label{ek}
\end{equation}
This is the same expression as Dashen \cite{dashen} found at
zero temperature, except that instead of $f_\pi$, 
at nonzero temperature the real part of $f_\pi^t$ enters. 
A relation like (\ref{ek}) was obtained by Thorsson and
Wirzba \cite{gmort}; 
they did not recognize, however, that in general $f_\pi^t$
has an imaginary part, and so wrote just $f_\pi^t$ instead of
$\mbox{Re} \, f_\pi^t$.

The pion mass in (\ref{ek}) is the dynamic pion mass, defined 
as the position of the singularity in the pion propagator in the
complex $p_0$ plane at $p=0$.  Alternately, we can introduce
the static pion mass, as the position of the singularity
in the pion propagator for $p_0 = 0$ in the complex $p$ plane.  From 
the form of the pion propagator,
$m_\pi^{static} = m_\pi/v$, and so by 
(\ref{ed}) and (\ref{ek}) this is just
\begin{equation}
(m^{static}_\pi)^2 = 
\frac{2 m \langle \overline{q} q\rangle_T}{\mbox{Re} f_\pi^s  \; 
\mbox{Re}f_\pi^t} \; .
\label{el}
\end{equation}
Obviously, $v \leq 1$ implies that 
\begin{equation}
m_\pi^{static} \geq m_\pi \; .
\end{equation}

We now consider where these effects first appear in
an expansion about zero temperature.  Using either
a nonlinear \cite{gassert} or a linear \cite{boch}
sigma model, to leading order in $T^2/f_\pi^2$,
\begin{equation}
f_\pi^t(T) = f_\pi^s(T)
= \left(1 - {T^2\over12 f_\pi^2}\right) f_\pi \; .
\end{equation}
Hence to leading order in low temperature, pions move
at the speed of light and are undamped.
This was established by
Dey, Eletsky, and Ioffe \cite{dey},
who showed that to $\sim T^2/f_\pi^2$, 
the thermal average of the two point function of
either vector or axial vector currents is directly proportional
to a linear combination of those at zero temperature.  Since
these two point functions are lorentz covariant at zero temperature,
they remain so to $\sim T^2/f_\pi^2$.

Thus the first place where the effects which we are discussing can enter
is at next to leading order, $\sim T^4$ \cite{dyson}.
The pion damping rate \cite{non2} and self energy
\cite{schenka,schenkb} have been computed to $\sim T^4/f_\pi^4$
in a nonlinear sigma model .
In particular, Schenk \cite{schenka,schenkb}
computed the pion self energy 
not in the chiral limit, but using physically reasonable
approximations.  His results imply that
for $T = 150 MeV$, $v \sim .87$ \cite{dip}.

Instead of computing to two loop order in a nonlinear sigma model,
to illustrate the effect we calculate, in weak coupling,
to one loop order in a linear sigma model.
In euclidean space time the lagrangian is
\begin{equation}
{\cal L} = \frac{1}{2} \left( \partial_\mu \phi \right)^2
- \frac{\mu^2}{2} \phi^2 
+ \frac{\lambda}{4} \left(\phi^2 \right)^2 - h \sigma \; ,
\label{eo}
\end{equation}
where $\phi = (\sigma,\vec{\pi})$ is an $O(4)$ isovector field.
We introduce a background magnetic field $h$ which is proportional
to the current quark mass $m$.  For $h=0$,
the vacuum expectation value of the $\sigma$ is 
$\sigma_0 = \sqrt{\mu^2/\lambda}$, where we then shift
$\sigma \rightarrow \sigma_0 + \sigma$; 
for two flavors, $f_\pi = \sigma_0$.

We compute terms of order $\sim T^4/(f_\pi^2 m_\sigma^2)$.  There
are many terms of order $\sim T^4/f_\pi^4$, at both one and two loop
order.  In weak coupling, however, 
the $\sigma$ meson is light relative to $f_\pi$,
$m_\sigma^2 = 2 \lambda f_\pi^2$.
Thus in weak coupling, which we assume, the terms of order 
$\sim T^4/f_\pi^4$ are smaller by $\sim \lambda$
than those computed.
Conversely, in the limit of strong coupling,
$m_\sigma \rightarrow \infty$, 
the only terms are those $\sim T^4/f_\pi^4$.

The diagrams for the pion self energy have been computed to 
$\sim T^4/(f_\pi^2 m_\sigma^2)$
by Itoyama and Mueller \cite{im}; 
$f^t_\pi = f^s_\pi$
has been computed to $\sim T^2/f_\pi^2$ 
by Bochkarev and Kapusta \cite{boch}, so all
we have to do is extend the calculation of $f^t_\pi$ and $f^s_\pi$
to this order.  Consequently, we
merely sketch the simplest way of performing the calculations.
For the real parts, the terms of interest
arise from diagrams involving a
virtual $\sigma$ and a $\pi$ in a loop, such as 
\begin{equation}
{\cal I}(P) = tr_K \frac{1}{K^2 ((P-K)^2 + m^2_\sigma) } \; ,
\label{ep}
\end{equation}
where $tr_K = T \sum_{n = -\infty}^{+\infty} \int d^3 k/(2 \pi)^3$.
To $\sim T^2$, it suffices to approximate this integral by
its value at zero momentum, neglecting the $K$ dependence in the
$\sigma$ propagator, so (\ref{ep}) becomes
\begin{equation}
{\cal I}(P) \sim 
\frac{1}{m_\sigma^2} tr_K \frac{1}{K^2} \sim 
\frac{1}{m_\sigma^2} \frac{T^2}{12} \; .
\label{eq}
\end{equation}
In the integral we have ignored apparent ultraviolet divergences
to concentrate on the term $\sim T^2$.  Of course renormalization
is taken care of as usual at zero temperature.

To compute terms of $\sim T^4$, it is necessary to expand the
integral in (\ref{ep}) to $\sim P^2$, 
including both terms $\sim P^2$ and terms
$\sim P^\mu P^\nu$.  Besides the integral in (\ref{eq}), we
also need
\begin{equation}
tr_K \frac{K^\mu K^\nu}{K^2} \sim 
\left( \delta^{\mu \nu} - 4 n^\mu n^\nu \right)
\frac{\pi^2 T^4}{90} \; ,
\label{er}
\end{equation}
where $n^\mu = (1,\vec{0})$.

The imaginary part of expressions cannot be extracted so
easily.  We evaluate the imaginary part only near the
pion mass shell, which is for $\omega \sim p$.  In
this region, the only contribution to the imaginary part of 
(\ref{eq}) is from
\begin{equation}
\mbox{Im} \; {\cal I}(P) =
\int \frac{d^3 k}{(2 \pi)^3}
\frac{\pi( n_1 - n_2)}{4 E_1 E_2}
\delta \left( \omega + E_1 - E_2 \right) \; .
\label{es}
\end{equation}
In this expression $E_1 = k$ is the energy of the
pion, $E_2 = \sqrt{(p-k)^2 + m_\sigma^2}$ is the
energy of the $\sigma$, and $n_1 = n(E_1)$,
$n_2 = n(E_2)$ are the corresponding
Bose-Einstein distribution functions.
This result can be obtained in various ways, such
as following \cite{rp}.  In all there are
four possible $\delta$-functions in energy which
contribute to $\mbox{Im} \; {\cal I}(P)$.  For 
$\omega \sim p \ll T$
only that in (\ref{es}) contributes, and corresponds
to Landau damping.
In this region, the $\delta$-function requires
\begin{equation}
k = \frac{m^2_\sigma}{2 (\omega + p \, cos\theta)} \; .
\label{et}
\end{equation}
We assume that $k \gg m_\sigma$, and then
expand the energies accordingly; 
this is justified, since from (\ref{et}), when $m_\sigma~\gg~\omega,p$, 
then $k~\gg~m_\sigma$.  The result for the imaginary part is
\begin{equation}
\mbox{Im} \; {\cal I}(P)|_{\omega \sim p \ll m_\sigma}
\sim \frac{1}{16 \pi} \; 
exp \left( - \frac{m^2_\sigma}{4 p T} \right) \; .
\label{eu}
\end{equation}
Because the fields being scattered have large
momentum, the Bose-Einstein distribution functions are
essentially Boltzman, which generates the exponential suppression
seen in (\ref{eu}).  

These integrals are sufficient to reproduce the results of
ref. \cite{im} for the pion self energy.
To evaluate the corresponding terms for the pion structure
constants, we need the axial current in the linear sigma
model,
\begin{equation}
A_\mu^{a} = (\sigma_0 + \sigma) \partial_\mu \pi^a 
- \pi^a \partial_\mu \sigma \; .
\label{ex}
\end{equation}
The diagrams which contribute at one loop order to 
$f^t_\pi$ and $f^s_\pi$ are given in fig. (5) of \cite{boch}.  Besides
the pion self energy, there is a contribution from a
$\sigma$-$\pi$ loop at the vertex for $A^\mu_{a}$.
These contributions can be evaluated expanding integrals
like (\ref{ep}) and using (\ref{er}).  For the imaginary
parts, we need the integrals
$$
tr_K \frac{k^0}{K^2 ((P-K)^2 + m_\sigma^2)}|_{\omega \sim p \ll m_\sigma}
$$
\begin{equation}
\sim \frac{i}{16 \pi} \left( \frac{m^2_\sigma}{4 p} + T \right)
\; exp \left( - \frac{m^2_\sigma}{4 p T} \right) \; .
\label{ey}
\end{equation}
and
$$
tr_K \frac{k^i}{K^2 ((P-K)^2 + m_\sigma^2)}|_{\omega \sim p \ll m_\sigma}
$$
\begin{equation}
\sim \frac{p^i}{16 p \pi} \left( \frac{m^2_\sigma}{4 p} - T \right)
\; exp \left( - \frac{m^2_\sigma}{4 p T} \right) \; .
\label{ez}
\end{equation}

The results of the computations are as follows.  At one
loop order the quantity
\begin{equation}
t_1 = \frac{T^2}{12 f_\pi^2} \;
\label{eza}
\end{equation}
typically enters.  To the order we work, we also need
$$
t_2 = \frac{\pi^2}{45} \frac{T^4}{f_\pi^2 m_\sigma^2} \; ,
$$
\begin{equation}
t_3 = \frac{1}{32 \pi} \frac{m^4_\sigma}{f_\pi^2 p^2}
\; exp \left( - \frac{m^2_\sigma}{4 p T} \right) \; .
\label{fa}
\end{equation}
Then at weak coupling in the linear sigma model,
to $\sim T^4/(f_\pi^2 m_\sigma^2)$,
$$
f_\pi^t
\sim  \left( 1 - t_1 + 3 t_2 + i t_3\right) f_\pi\; ,
$$
\begin{equation}
f_\pi^s
\sim \left( 1 - t_1 - 5 t_2 - i t_3\right) f_\pi \; .
\label{fb}
\end{equation}
By (\ref{ed}) the pion velocity is
\begin{equation}
v^2 \sim 1 - 8 t_2 \; ,
\label{fc}
\end{equation}
while from (\ref{eab}) the pion mass shell is 
\begin{equation}
i p^0 \sim v p - i p \, t_3 \; .
\label{fd}
\end{equation}
Including the one loop self energy computed to this order,
the pion propagator is
$$
Z_\pi \Delta^{-1}(P)|_{\omega \sim p \ll m_\sigma} \sim
(1 + t_1 + 6 t_2) p_0^2
+ (1 + t_1 -  2 t_2) p^2
$$
\begin{equation}
+ m^2_\pi \left( 1 + 3 t_1/2 \right) - 2 i p^2 t_3 \; .
\label{fe}
\end{equation}
To ensure that $\Delta^{-1}(P)$
has canonical normalization
we introduce a factor for wave function 
renormalization of the pion,
\begin{equation}
Z_\pi \sim 1 + t_1 + 6 t_2 \; .
\end{equation}

It is elementary to check that the zero of (\ref{fe}) agrees
with (\ref{fd}).  We have included the results to leading order
in the external field $h$, when the pion mass is nonzero.
Assuming that the quark condensate is proportional
to the vacuum expectation value of the $\sigma$ field,
\begin{equation}
\langle \overline{q} q \rangle_T \sim \sigma_0(T) 
\sim \sigma_0(0) \left( 1 - 3 t_1/2 \right) \; ,
\label{ff}
\end{equation}
we also verify our generalization of the formula of
Gell-Mann, Oakes, and Renner in (\ref{ek}) for
the dynamic pion mass:
\begin{equation}
m^2_\pi(T)  \sim m^2_\pi \left(
1 + t_1/2 - 6 t_2 \right) \; .
\label{fg}
\end{equation}

We conclude with some general comments.  First,
while the effects computed at low temperature
(\ref{eza}) - (\ref{fg}) are small, that does
not mean that they remain so for temperatures of physical
interest, as seen in the results of
\cite{pioncon}-\cite{nucl}.
Secondly, the form of the inverse propagator
in (\ref{eh}) applies
not just to Goldstone bosons, but to 
any scalar field at nonzero temperature.  For example,
in numerical simulations 
on the lattice in euclidean spacetime, typically
what is measured is only the static mass, not the dynamic.

Finally, we note that the coefficient of $v^2 -1$ in
(\ref{fc}) is proportional to the free energy density for pions,
$= \pi^2 T^4/30$.  (It would be interesting to know what
the analogous coefficient is for the nonlinear sigma model in the
chiral limit.)
This and other examples \cite{adler} hint of a general
relation, valid for all temperatures, where the deviation
of the velocity squared from unity is proportional to the free energy
density \cite{rpmt}.

We thank R. Brout for enlightenment on ferromagnets, 
G. Castillo and S. Chakravarty for the same on antiferromagnets, 
K. Scharnhorst for bringing ref. \cite{adler} to our attention,
A. Schenk for discussions on his work,
and M. Creutz for comments.
This work is supported by a DOE grant at 
Brookhaven National Laboratory, DE-AC02-76CH00016.

\end{narrowtext}

\begin{references}
\bibitem{gold}
Y. Nambu, Phys. Rev. Lett. {\bf 4}, 380 (1960);
J. Goldstone, Nuevo Cimento {\bf 19}, 154 (1961);  
J. Goldstone, A. Salam and S. Weinberg, Phys. Rev. {\bf 127}, 965 (1962).
%
\bibitem{donoghue}
J. F. Donoghue, E. Golowich, and B. R. Holstein, 
{\it Dynamics of the Standard Model} (Cambridge Univ. Press, NY, 1992).
%
\bibitem{sv}
G. M. Shore and G. Veneziano, Nuc. Phys. {\bf B381}, 23 (1992).
%
\bibitem{gmor}
M. Gell-Mann, R. J. Oakes and B. Renner, Phys. Rev. {\bf 175}, 2195 (1968).
%
\bibitem{dashen}
R. Dashen, Phys. Rev. {\bf 183}, 1245 (1969); 
Phys. Rev. {\bf D3}, 1879 (1971).
%
\bibitem{rpmt}
For a pedagogical discussion of our results, see
R. D. Pisarski and M. Tytgat, 
to appear in {\it Continuous Advances in $QCD$ '96}, 
World Scientific Publishing, hep-ph/9606459.
%
\bibitem{gmort}
V. Thorsson and A. Wirzba, Nucl. Phys. {\bf A589}, 633 (1995).
%
\bibitem{pioncon}
G. E. Brown and W. Weise, Phys. Rep. {\bf 27}, 1 (1976);
A. B. Migdal, Rev. of Mod. Phys. {\bf 50}, 107 (1978); 
K. Kolehmainen and G. Baym, Nuc. Phys. {\bf A382}, 528 (1982).
%
\bibitem{shuryak}
E. V. Shuryak, Phys. Rev. {\bf D42}, 1764 (1990),
Nucl. Phys. {\bf A533}, 761 (1991);
E. Shuryak and V. Thorsson, Nucl. Phys. {\bf A536}, 739 (1992).
%
\bibitem{nucl}
C. Gale and J. Kapusta, Phys. Rev. {\bf C35}, 2107 (1987);
Phys. Rev. {\bf C38}, 2659 (1988); for recent results, see
G. Q. Li and C. M. Ko, Nucl. Phys. {\bf A582}, 731 (1995);
C. Song, S. H. Lee, C. M. Ko, Phys. Rev. {\bf C52}, 476 (1995),
and references therein.
%
\bibitem{anti}
A. B. Harris, D. Kumar, B. I. Halperin, and P. C. Hohenberg,
Phys. Rev. {\bf B3}, 961 (1971).
%
\bibitem{leutnon}
H. Leutwyler, Phys. Rev. {\bf D49}, 3033 (1994).
%
\bibitem{im}
H. Itoyama and A. H. Mueller, Nuc. Phys. {\bf B218}, 349 (1983).
%
\bibitem{kr}
M. Kirchbach and D. O. Riska, Nucl. Phys. {\bf A578}, 511 (1994).
%
\bibitem{fpi}
The theory of critical phenomena tells us that the vacuum expectation
vanishes like $\sim (T_\chi - T)^\beta$, where $\beta \sim .38$ for
an $O(4)$ sigma model.
We expect that $f_\pi^t$, $f_\pi^s$, and 
$\langle\overline{q} q\rangle_T$ all vanish in this way.
From (\ref{ei}), then, the ratio between 
the quark operator and the pion field, (\ref{ef}), is finite
and nonzero at $T_\chi$.
%
\bibitem{non2}
S. Gavin, Nucl. Phys. {\bf A435}, 826 (1985);
J. L. Goity and H. Leutwyler, Phys. Lett. {\bf B228}, 517 (1989);
P. Gerber, H. Leutwyler, J. L. Goity,
Phys. Lett. {\bf B246}, 513 (1990). 
%
\bibitem{schenka}
A. Schenk, Nucl. Phys. {\bf B363}, 97 (1991).
%
\bibitem{schenkb}
A. Schenk, Phys. Rev. {\bf D47}, 5138 (1993).
%
\bibitem{antidamp}
For spin waves in antiferromagnets,
$\gamma \sim p^2$ as $p \rightarrow 0$, \cite{anti}. 
%
\bibitem{gassert}
J. Gasser and H. Leutwyler, Phys. Lett. {\bf 184B}, 83, 1987.
%
\bibitem{boch}
A. Bochkarev and J. Kapusta, Univ. of Minn. preprint
NUC-MINN-95-25-T, Dec. 1995, hep-ph/9602405.
%
%
\bibitem{dey}
M. Dey, V. L. Eletsky, and B. L. Ioffe, Phys. Lett. {\bf 252B}, 620 (1990);
V. L. Eletsky and B. L. Ioffe,  Phys. Rev. {\bf D47}, 3083 (1993);
Phys. Rev. {\bf D51}, 2371 (1995);
Z. Huang, Phys. Lett. {\bf B361}, 131 (1995).
%
\bibitem{dyson}
This is analogous to the behavior of spin waves in 
ferromagnets, where the dispersion relation is
$\omega(p) = c(T) p^2$.  
The coefficient $c(T)$ changes with temperature, but
as shown first by Dyson, in an expansion
about zero temperature the corrections begin not at
leading, but only at next to leading order.
See: F.J. Dyson, Phys. Rev. {\bf 102}, 1217 (1956);
R.B. Stinchcombe, G. Horwitz, F. Englert and 
R. Brout, Phys. Rev. {\bf 130}, 155 (1963).
%
\bibitem{dip}
In  fig. 5 of ref. \cite{schenka} and fig. 7 of ref. \cite{schenkb},
Schenk plots
$R(p)$, the ratio of the quasiparticle energy, to the pion energy
in free space, as a function of momentum.  
To two loop order, as $p$ increases from zero there is a ``dip'' in $R(p)$:
it first decreases, and then increases,
approaching one from
below.  This is only possible if the quasiparticle energy
$\omega(p)^2 = v^2(p)  p^2 + m_\pi^2(T)$,
with $v(0)^2 < (m_\pi(T)/m_\pi(0))^2 < 1$ \cite{rpmt}.  
A. Schenk, private communication, estimates that 
$v(0) \sim .87$.  Our analysis only applies to the
pion quasiparticle energy about zero momentum, and so $v(0)$.
%
\bibitem{rp}
R. D. Pisarski, Nucl. Phys. {\bf B309}, 476 (1987).
%
\bibitem{adler}
This is analogous to the
propagation of light in a background magnetic field,
where at two loop order, the velocity squared is less than
one by an amount proportional to the energy density,
S.L. Adler, Ann. Phys. 67, 599 (1971).
A similar result holds for light propagating 
in a thermal bath, 
J.I. Latorre, P. Pascual and R. Tarrach, Nucl. Phys. {\bf B437}, 60 (1995).
The latter analogy is less precise, because due to
Debye screening, thermalized
photons have a mass gap at zero spatial momentum.
\end{references}
\end{document}